\documentclass{aa}  

\usepackage{graphicx}
\usepackage{txfonts}
\usepackage{threeparttable}
\usepackage{longtable, tabu}
\usepackage{lscape}
\usepackage{hyperref}
\usepackage{orcidlink}
\usepackage[normalem]{ulem}
\usepackage{natbib}
\bibpunct{(}{)}{;}{a}{}{,} 
\begin{document}

   \title{X-ray counterparts to stellar MeerKAT Galactic-plane compact radio sources}


   \author{Okwudili D. Egbo \orcidlink{0009-0001-0232-3968}
          \inst{\ref{inst1},}
          \inst{\ref{inst2}}
          \thanks{Corresponding author; egbo@saao.ac.za 
         }
          \and
          P. J. Groot \orcidlink{0000-0002-4488-726X}
          \inst{\ref{inst1},}
          \inst{\ref{inst2},}
          \inst{\ref{inst3}}
          \and
          D. A. H. Buckley \orcidlink{0000-0002-7004-9956}
          \inst{\ref{inst1},}
          \inst{\ref{inst2},}
          \inst{\ref{inst4}}
          \and
          J. Robrade \orcidlink{0000-0002-4939-8940}
          \inst{\ref{inst5}}
          \and
          A. D. Schwope \orcidlink{0000-0003-3441-9355}
          \inst{\ref{inst6}}
          \and
          S. Freund \orcidlink{0009-0006-3658-4935}
          \inst{\ref{inst5},}
          \inst{\ref{inst7}}
          \and
          P. C. Schneider \orcidlink{0000-0002-5094-2245}
          \inst{\ref{inst5}}  
          \and
          B. Stelzer 
          \inst{\ref{inst8}} 
          }

   \institute{Department of Astronomy, University of Cape Town, Private Bag X3, Rondebosch 7701, South Africa\label{inst1}     
         \and
             South African Astronomical Observatory, PO Box 9, Observatory 7935, South Africa\label{inst2}
        \and
             Department of Astrophysics/IMAPP, Radboud University, PO 9010,  NL-6500 GL Nijmegen, the Netherlands\label{inst3}
        \and
             Department of Physics, University of the Free State, PO Box 339, Bloemfontein 9300, South Africa\label{inst4}
        \and
             Hamburger Sternwarte, University of Hamburg, Gojenbergsweg 112, 21029 Hamburg, Germany\label{inst5}
        \and
             Leibniz-Institut f\"ur Astrophysik Potsdam (AIP), An der Sternwarte 16, 14482 Potsdam, Germany\label{inst6}
        \and
             Max-Planck-Institut f\"ur extraterrestrische Physik, Gie\ss enbachstra\ss e 1, 85748 Garching, Germany\label{inst7}
        \and
            Institut f\"ur Astronomie Astrophysik, Eberhard Karls Universit\"at T\"ubingen, Sand 1, 72076 T\"ubingen, Germany\label{inst8}
             }

   \date{Received September 15, 1996; accepted March 16, 1997}

  \abstract
    {Radio emission from magnetically active stars arises primarily from non-thermal processes and provides information complementary to their high-energy X-ray emission. Recent advances in sensitive, wide-field radio and X-ray surveys have enabled the identification of larger samples of active stars across the Galaxy.}
    {We aim to identify and characterise radio and X-ray-emitting stars in the Galactic plane by combining MeerKAT radio data with soft X-ray observations, and to assess their consistency with the canonical G\"{u}del-Benz relation, which states that the thermal coronal emission observed in soft X-rays is correlated with the non-thermal gyrosynchrotron emission detected in the radio.}
    {We cross-matched compact radio sources from the SARAO MeerKAT Galactic Plane Survey with soft X-ray counterparts from the {\it ROSAT} All-Sky Survey and the first release of the {\it SRG/eROSITA} all-sky survey, the eRASS1. We investigated their radio-brightness temperatures and computed radio and X-ray luminosities to test for consistency with the G\"{u}del--Benz relation.}
    {We identify 137 stellar sources with both radio and X-ray detections. Their radio-brightness temperatures, {\bf $T_B$}, range from $10^7$ to $10^{12}$ K, with the exception of two outliers: AXJ1600.9--5142 ($T_B = 4.8 \pm 1.5 \times 10^{12}$~K) and HD~124831 ($T_B = 8 \pm 1 \times 10^{6}$~K). The remaining sources are consistent with emission from incoherent gyrosynchrotron processes. The sample generally lies below the canonical G\"udel--Benz relation, with the offset predominantly driven by enhanced radio luminosities at 1.3~GHz relative to the canonical relation derived at 5~GHz.}
    {Our results suggest that the classical G\"{u}del--Benz relation represents an upper envelope rather than a tight correlation for active stars detected at 1.3 GHz. In addition, the {\it eROSITA} detections show that early-type stars are systematically below the $\log (L_{\rm X}/L_{\rm bol}) \sim -3$ relation, which is usually seen for stars of a later spectral type. 
    }


   \maketitle
%

\section{Introduction}\label{sec:intro}

    Magnetic activity is a common phenomenon associated with late-type stars and is driven by a magnetohydrodynamic (MHD) dynamo mechanism arising from the motion of plasma in the convective outer layers of the star \citep{Schrijver:1984, Schrijver:1985, Nandy:2004, Forbrich:2011, Testa:2015}. Observationally, magnetic activity is associated with events such as stellar flares and coronal mass ejections \citep{Noyes:1984, Schrijver:1985}. 
    Factors that influence the strength and time scales of stellar magnetic activity include the rotation rates, stellar masses, and spectral types \citep{Skumanich:1975, Noyes:1984}. Magnetically active systems are of great interest because they allow for the study of both thermal and non-thermal emission processes in the outer atmospheres of solar-type stars. For the thermal case, chromospheric activity manifests itself spectroscopically in Ca\,{\sc ii} (core) line emission, which is most evident in the Ca\,{\sc ii} H and K lines \citep{Skumanich:1975, Schrijver:1985}. 
    At X-ray wavelengths, the stellar emission is predominantly produced by hot, optically thin plasma associated with magnetically confined coronae in late-type stars or shock-heated winds in early-type stars \citep{Gudel:2004, Zhuleku:2020}. The resulting X-ray spectra are dominated by strong, narrow emission lines from highly ionised elements such as iron, oxygen, neon, and magnesium \citep{Gudel:2004, Testa:2010a, Testa:2010b}. These lines provide key diagnostics of plasma temperature, density, and chemical composition. While continuum processes such as thermal bremsstrahlung and recombination (free-free and free-bound emission) contribute to the overall flux, their role is generally minor \citep{Gudel:2004, Zhuleku:2020}. In low or moderate spectral resolution observations, the dense line forest can appear as a quasi-continuum \citep{Franciosini:1995}.

    Magnetic activity is also seen in the radio, where it is linked to non-thermal emission \citep{Dulk:1985, Drake:1987, Hjellming:1988, Morris:1990, Linsky:1996}, and this emission can be coherent or incoherent \citep{Dulk:1985, Gudel:2002}. Coherent emission is usually driven by electron cyclotron maser instabilities (ECMI) or by plasma instabilities \citep{Toet:2021, Vedantham:2020}.
    Anisotropic distributions of energetic electrons in a dense thermal plasma can excite Langmuir waves, which may convert into coherent radio emission near the plasma frequency, resulting in a radio plasma emission mechanism \citep{Vedantham:2020}. For ECMI, electrons moving in a magnetic field emit at low radio frequencies \citep{Yiu:2024}. On the other hand, incoherent emission is mostly continuum emission arising from three mechanisms: thermal bremsstrahlung, gyroresonance, and gyrosynchrotron \citep{Dulk:1985, Gudel:2002,  Trigilio:2018}.

    \cite{Guedel:1993} first established a relationship between radio and X-ray emission for late-type stars, specifically low-mass, chromospherically active stars of spectral types F to M. Among these, the most X-ray luminous objects are frequently RS CVn, BY Dra, or Algol-type binaries. Using specific radio luminosities, $L_{\nu,R}$, between 5 and 9 GHz, as well as {\sl R\"ontgen Satellite (ROSAT)} soft X-ray luminosities, $L_X$, in the 0.1 - 2.4 keV band, they established a relation between radio and X-ray luminosity ($L_X/L_{\nu,R}$), which is now commonly called the G\"{u}del--Benz relation (GBR). The X-ray luminosities range from $\sim 10^{27} - 10^{32}$ erg$\,$s$^{-1}$, whereas the specific radio luminosities range from $10^{12} - 10^{18}$ erg$\,$s$^{-1}\,$Hz$^{-1}$ \citep{Guedel:1993, Benz:1994, Vedantham:2022, Driessen:2022, Driessen:2024}. 
    The GBR is canonically written in the form $\frac{L_X}{L_{\nu,R}} = \kappa  10^{15.5\pm 0.5}$, with $\kappa \leq 1$, depending on the type of stars \citep{Diaz-Marquez:2024}. \cite{Williams:2014} showed that the GBR holds over ten orders of magnitude in radio spectral luminosity, from main-sequence close binaries, such as RS CVn systems, down to solar nano-flares, and is described in \cite{Williams:2014} as
    \begin{equation}
        L_X = 9.48 \cdot 10^{18} \, L_{\nu, R}^{0.73}
        \label{gbreq2}
    ,\end{equation}
    where $L_X$ is the soft X-ray luminosity in erg$\,$s$^{-1}$ and $L_{\nu,\text{R}}$ is the radio spectral luminosity in erg$\,$s$^{-1}\,$Hz$^{-1}$.

    As more radio- and X-ray-emitting stars have been discovered, further investigations have been carried out to validate the GBR. \cite{Vedantham:2022} used the Low-Frequency Array (\textit{LOFAR}; \citealt{vanHaarlem:2013}) Two Meter Sky Survey (LoTSS; \citealt{Shimwell:2017, Shimwell:2019}), identified 24 active stars that mostly comprise RS CVn and BY Dra active binaries. The observations made at a low frequency of 144 MHz showed high circular polarisation fractions, and the GBR correlation still holds despite the emission being coherent. \cite{Launhardt:2022} identified 31 young active stars within 130 pc and younger than 200 Myr using the Karl G. Jansky Very Large Array (VLA) at a 6 GHz frequency. Almost all sources were also detected with \textit{ROSAT} and all follow the canonical GBR. \cite{Yiu:2024} using LoTSS and the VLA Sky Survey (VLASS; \citealt{Lacy:2020, Gordon:2021}) further showed the GBR relationship to hold, using binaries and single stars. The VLASS samples also follow the GBR, though there are some deviators, particularly M dwarfs of a spectral type later than M4. \cite{Driessen:2024} used the Sydney Radio Star Catalogue (SRSC) to further show that the relationship holds for circularly polarised detections made with the Australia Square Kilometre Array Pathfinder (ASKAP), combined with X-ray data from the \textit{\emph{extended R\"ontgen Survey with an Imaging Telescope Array} (eROSITA}, \citealt{Predehl:2021}). 

    In this work, we used radio data from the South African Radio Astronomical Observatory (SARAO) MeerKAT Galactic Plane Survey (SMGPS; \cite{Goedhart:2024}) at 1.3 GHz alongside \textit{ROSAT} \citep{Truemper:1982, Boller:2016, Freund:2022} and \textit{eROSITA} \citep{Predehl:2021, Merloni:2024, Freund:2024} soft X-ray data to investigate the GBR. Detailed information about the radio and X-ray surveys is described in Section \ref{sec:data}, and the methodology involved in identifying radio- and X-ray-emitting stars and additional analysis is described in Section \ref{sec:method}. We present our findings and discuss the implications of our results in Section \ref{sec:result}.

\section{Data description} \label{sec:data}
    
    \subsection{SARAO MeerKAT Galactic Plane Survey compact source catalogue} \label{smgpsga}

    The South African Radio Astronomical Observatory (SARAO) MeerKAT Galactic Plane Survey (SMGPS) is a 1.3 GHz continuum survey of the Galactic plane between $251^\circ \leq l \leq 358^\circ$ and $2^\circ \leq l \leq 61^\circ$ at $|b| \leq 1.5^\circ$ \citep{Goedhart:2024}.
    Because of the Galactic-plane warp, the region with the Galactic longitude $251^{\circ} \leq l \leq 300^{\circ}$ was observed with an adjusted Galactic latitude range of $-2.0^{\circ} < b < 1.0^{\circ}$. SMGPS is the largest, most sensitive, and highest angular-resolution 1.3 GHz survey of the Galactic plane carried out to date, with an angular resolution of $ 8\arcsec$, source positions better than $2\arcsec$ for 90\% of the sources, and a broadband sensitivity of $\sim$ 10-20 \, $\mu\text{Jy beam}^{-1}$. At 531 square degrees, it covers 1.2\% of the sky, limiting the chances of including rare or specific solar-neighbourhood sources. SMGPS observations were made between 2018 July 21 and 2020 March 14 using the $L$-band receiver system, covering a frequency range of 0.856 -- 1.712 GHz with 4096 channels, and an eight-second correlator integration period \citep{Goedhart:2024}. 

    This work used the compact source catalogue derived from SMGPS by \cite{Mutale:2025} using the \textsc{Aegean} source finder \citep{Hancock:2018}. The catalogue comprises 443,455 sources with a signal-to-noise, (S/N) of $\geq$5, and the source positions are better than $2\arcsec$ for 90\% of the sources.  Of the 443,455 compact sources, \cite{Egbo:2025} reported 629 {\it Gaia}-detected stars as the optical counterparts to stellar radio sources in the SMGPS compact source catalogue.

    \subsection{X-ray catalogues}
    
    \subsubsection{{\it ROSAT}}\label{rosatsect}
    
    The {\sl R\"ontgen Satellite (ROSAT)} was an X-ray observatory developed by a collaboration of Germany, the United States, and the United Kingdom. It was launched in 1990 to study soft X-ray emission from astronomical sources \citep{Truemper:1982}.
    The \textit{ROSAT} All-Sky Survey (RASS) is a soft X-ray survey performed with the Position-Sensitive Proportional Counter (PSPC) on board \textit{ROSAT} between June 1990 and August 1991. It was the first comprehensive soft X-ray all-sky survey in the energy range of 0.1--2.4 keV \citep{Voges:1999, Boller:2016}.
    The catalogue includes more than 135,000 point-source detections and is referred to as the Second ROSAT all-sky survey (2RXS) catalogue \citep{Boller:2016}.  
    The detection limit of the catalogue is not uniform across the sky, as it depends on variations in exposure time and background noise. For stellar sources, the sensitivity spans a range from $3 \times 10^{-14}$ to $5 \times 10^{-13}\, \mathrm{erg \, s^{-1} \, cm^{-2}}$, with 99\% of 2RXS sources falling within this interval with a mean flux limit of $\approx 1.5 \times 10^{-13} \, \mathrm{erg \, s^{-1} \, cm^{-2}}$ \citep{Boller:2016, Freund:2022}. 
    RASS sources have a systematic error of at least $1\arcsec$, increasing to as much as $20.4 \arcsec$ in the worst cases \citep{Boller:2016}. 
    To convert RASS count rates to energy fluxes, the survey uses the fitted conversion factor for the hardness ratio (HR), $\mathrm{CF} = (8.31 + 5.30HR) \cdot 10^{-12}\, \mathrm{erg \, cm^{-2}}$ per count, based on a two-component Raymond--Smith coronal plasma model \citep{Schmitt:1995, Fleming:1995}. This same conversion method was adopted by \cite{Freund:2022} and \cite{Vedantham:2022} in their analyses of coronal X-ray sources.

    \subsubsection{{\it eROSITA}}\label{erasssect}

    \textit{eROSITA} \citep{Predehl:2021} is the soft X-ray instrument onboard the Russian Spectrum-R\"ontgen-Gamma (SRG) space observatory and was launched in July 2019. It operates in the 0.2-10.0 keV band and consists of seven identical telescope modules, each containing 54 nested mirror shells, and is equipped with its own CCD detector. The {\sl eROSITA} all-sky survey (eRASS) began in December 2019, and the first all-sky coverage, called eRASS1, was completed in June 2020.
    
    The main all-sky survey catalogue is produced in the 0.2--2.3 keV soft X-ray band; in addition, a hard catalogue (2.3--5.0  keV) is generated. The eRASS sensitivity depends on ecliptical scans on the sky-location, the average flux limit of eRASS1 is $\approx 5 \cdot 10^{-14} \, \mathrm{erg \, s^{-1} \, cm^{-2}}$ for the main catalogue.
    The eRASS1 catalogue of the western Galactic hemisphere, containing over 930\,000 detections, was released in January 2024 by the German {\sl eROSITA} Consortium as part of the DR1 data release \citep{Merloni:2024}. Positional uncertainties in eRASS1 range from $1\arcsec$ to $15\arcsec$, with an average $1\sigma$ uncertainty of $4.6\arcsec$ \citep{Freund:2024}. Count rates were converted to energy fluxes using an energy conversion factor (ECF) of $8.5 \times 10^{-13}\,\mathrm{erg\,cm^{-2}\,count^{-1}}$, derived from an APEC thermal plasma model appropriate for eRASS1 stellar coronae \citep{Freund:2024}. Because most stars in our sample are relatively nearby, the expected interstellar absorption and corresponding hydrogen column densities are low. For column densities up to a few times $10^{20}\,\mathrm{cm^{-2}}$, the resulting 0.2--2.3\,keV observed fluxes remain accurate within the measurement uncertainties (for example, a 1\,keV plasma retains $\sim 95\%$ of its flux at $N_{\mathrm{H}} = 10^{20}\,\mathrm{cm^{-2}}$). However, most eRASS1 detections lie very close to the survey's sensitivity limit and contain only a few source counts, making it generally impossible to derive reliable unabsorbed (absorption-corrected) fluxes from the X-ray data alone. For this reason, we report observed fluxes throughout.

    \subsubsection{{\it HamStar}}
    
    To characterise coronal sources from X-ray surveys, some members of the German {\it eROSITA} Consortium developed the {\sc HamStar} algorithm. {\sc HamStar} is a Bayesian method which searches for optical \textit{Gaia} counterparts within a search radius of $5\sigma$ of the positional uncertainty of the eRASS1 source \citep{Schneider:2022, Freund:2022}. It further combines the X-ray to \textit{Gaia} $G$-band flux ratio $\frac{F_X}{F_G}$, the \textit{Gaia} colour $BP-RP$, and the distance, $d$, of the counterpart to estimate the probability that the source is the correct counterpart and a coronal emitter \citep{Schneider:2022, Freund:2022, Freund:2024}. This method has been tested with several X-ray surveys to identify coronal emitters, e.g., \cite{Schneider:2022} used it to identify coronal emitters from the {\textit eROSITA }Final Equatorial-Depth Survey (eFEDS); \cite{Freund:2022} used it to identify stellar coronal emitters from the \textit{ROSAT} all-sky survey. From the performed cross-match between 2RXS and \textit{Gaia} Data Release 3 (DR3; see \citealt{GaiaCollaboration:2021}), they derived the stellar content of the RASS catalogue, comprising 28,630 sources (24.9\% of the total 2RXS survey sample).    Recently, the same method was used to characterise coronal emitters from half-sky release of eRASS1, where 138,800 sources were identified, about 15\% of the deeper survey sample from eROSITA \citep{Freund:2024}.

\section{Methodology} \label{sec:method}

    \subsection{Radio-X-ray cross-match}

     This work made use of the soft X-ray RASS (0.1 - 2.4 keV band) stellar content by \cite{Freund:2022} and the eRASS1 (0.2 - 2.3 keV band) coronal catalogue by \cite{Freund:2024} to search for X-ray stellar candidates in SMGPS radio sources. We accessed the \cite{Freund:2022} RASS catalogue and the \cite{Freund:2024} eRASS1 catalogue through the CDS VizieR online database \citep{Ochsenbein:2000}, including their \textit{Gaia} DR3 counterparts. Using the {\it Gaia} IDs of the 629 SMGPS--{\it Gaia} stars described in Section \ref{smgpsga}, we searched the RASS/eRASS1 coronal samples for matching {\it Gaia} IDs to identify radio--X-ray stellar counterparts. With a final matching radius of $r = 3\arcsec$, we estimated the number of spurious associations by performing Monte Carlo simulations in which the X-ray positions were randomly shifted in Galactic longitude while preserving their latitude distribution (see Section 3 of \cite{Egbo:2025} for details). Using this procedure, the expected number of false positives is $\sim6$ for eRASS1 and $\sim2$ for RASS, corresponding to false positive fractions of $\sim5\%$ for both catalogues. It confirms that contamination from chance coincidences in our matched samples is negligible.

    \subsection{Radio-brightness temperature}\label{btp}
    
    The radio-brightness temperature ($T_{\rm B}$) is generally used to infer whether radio emission from stellar sources is coherent or incoherent. We calculated the brightness temperature, $T_{\rm B}$ (in Kelvin), for each source using the MeerKAT radio flux as 
    \begin{eqnarray}
    T_\mathrm{ B} = \frac{S_{\text{1.3}} c^2}{2\pi k_\mathrm{ B}\nu ^2}\frac{D^2}{R_\mathrm{ e}^2},
    \label{btemp}
    \end{eqnarray}
    where $S_{\text{1.3}}$ is the flux density of the star at 1.3 GHz, $c$ is the speed of light (in metres per second), $k_{\rm B}$ is the Boltzmann constant (in Joules per Kelvin), $\nu$ is the frequency of observation (in Hertz, 1.3 GHz in our case), $D$ is the distance to the star (in metres), and $R_{\rm e}$ is the radius of the emitting region (in metres) \citep{Dulk:1985, Pritchard:2021}. Sources with $T_{\rm B} > 10^{12}\, \text{K}$ are considered coherent, whereas sources with $T_{\rm B} < 10^{12}\, \text{K}$ are considered incoherent \citep{Dulk:1985, Seaquist:1993}. Results from this analysis are presented in Section \ref{emmech}.

\section{Results and discussion} \label{sec:result}

    \subsection{Cross-matching results}

    \begin{figure}[ht]
    \centering
        \includegraphics[width=\columnwidth]{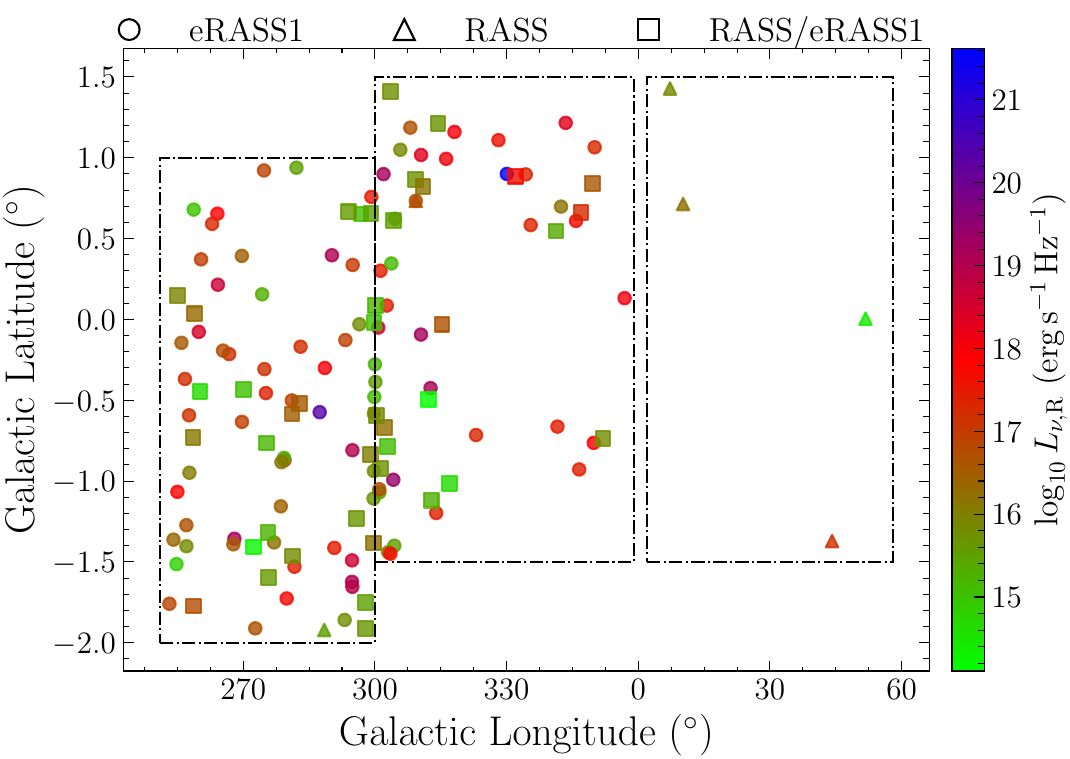}
        \caption{SMGPS and X-ray counterpart positions in Galactic coordinates. The circles, triangles, and squares represent eRASS1, RASS, and combined RASS and eRASS1 detections positions respectively. The rectangular dashed lines are SMGPS footprints.}
        \label{coords}
    \end{figure}
    
    We identify 137 objects with radio, optical, and X-ray detections. Of these, 47 X-ray identifications are from the RASS catalogue, whereas 131 are from eRASS1. For eRASS1, the 131 matches are expected to include only $\sim6$ random associations, while for RASS, the 47 matches are expected to include $\sim2$ spurious matches. There are 41 common detections between RASS and eRASS1. Of the six stars detected in RASS but not in eRASS1, four lie in the eastern part of the Galactic plane, where data rights belong to the Russian Consortium. For the remaining two, located in the western part of the plane, their non-detection in eRASS1 indicates flux variability.

\begin{figure}[ht]
    \centering
        \includegraphics[width=\columnwidth]{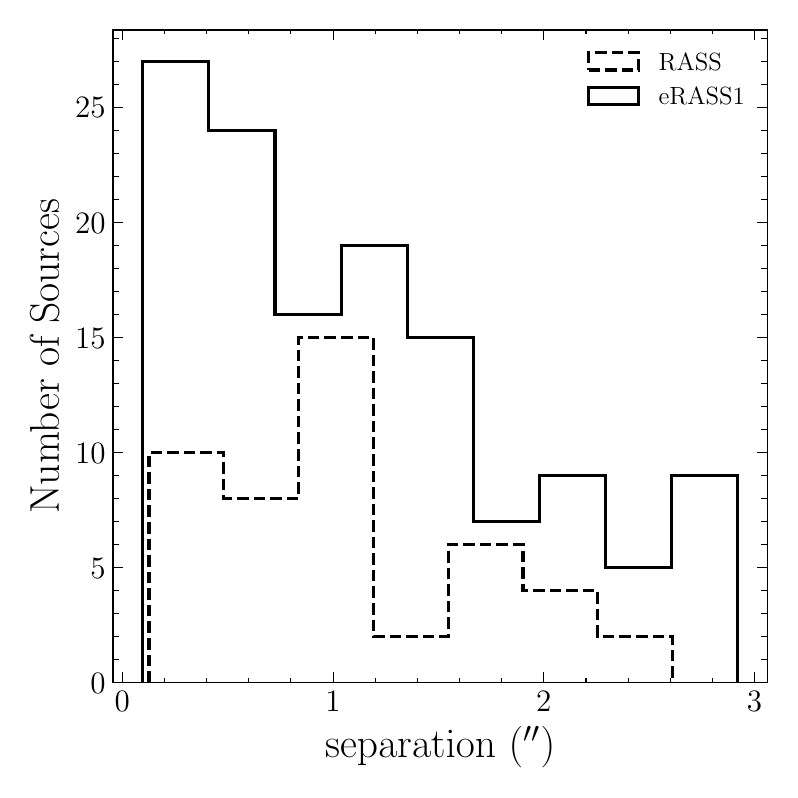}
        \caption{Positional offsets between SMGPS radio detections and RASS/eRASS1 X-ray counterparts. }
        \label{offset}
    \end{figure}

    The celestial positions of the detections are shown in Fig. \ref{coords}. All the eRASS1 detections are in the western Galactic plane due to the sky split between the German and Russian partners in SRG/{\sl eROSITA}. Of the 47 RASS matches, 25 are within $1 \arcsec$ of the nominal SMGPS positions, whereas the remaining 22 are at a $1\arcsec $--$ 3\arcsec$ offset. For the 131 eRASS1 sources, 66 are within a $1 \arcsec$ offset of the SMGPS positions. The positional offset between SMGPS and the RASS/eRASS1 sources is shown in Fig. \ref{offset}. The X-ray fluxes range from $2.07 \cdot 10^{-14} - 1.2 \cdot 10^{-11}$ \text{$ \rm erg \,s^{-1}\, cm^{-2}$}. It is important to highlight that these are observed/absorbed fluxes as described in Sects. \ref{rosatsect} and \ref{erasssect}. Our targets span a wide range of distances, from $\sim 20$ pc to $\sim 4.5$ kpc. For most sources, the available X-ray data do not contain sufficient counts to estimate the absorbing column density or to correct the measured X-ray fluxes for absorption. We therefore do not apply absorption corrections to the reported X-ray luminosities. Although $75\%$ of the sources are located within 1 kpc, absorption along the Galactic plane's line of sight can still be significant, and without sufficient photon statistics, it cannot be reliably accounted for.
    
\begin{figure}[ht]
    \centering
        \includegraphics[width=\columnwidth]{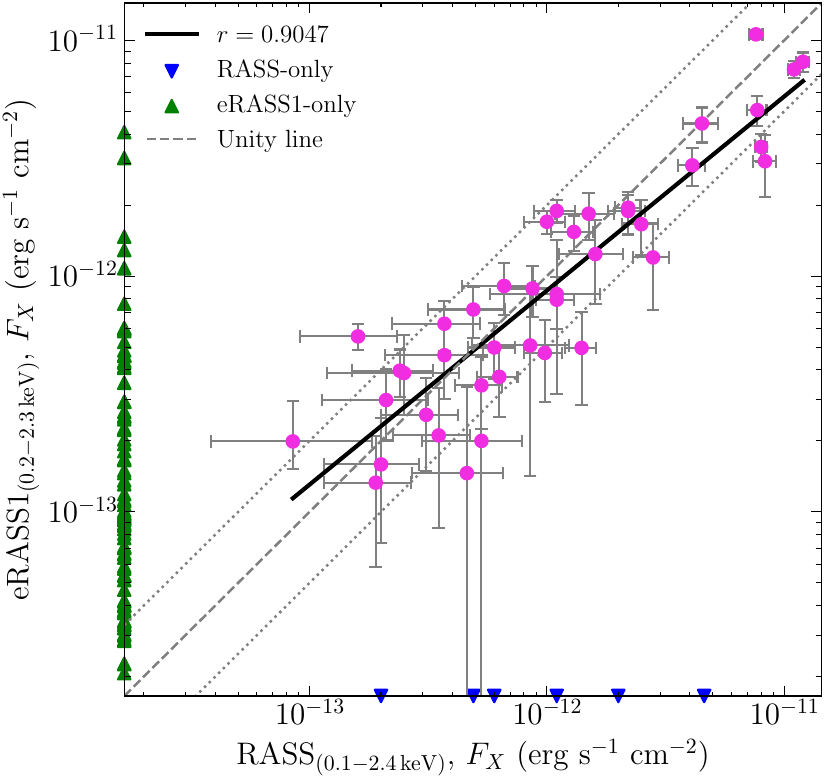}
        \caption{Comparison of X-ray fluxes from the 41 RASS (0.1 -- 2.4 keV band) and eRASS1 (0.2 -- 2.3 keV band) common detections. The solid line shows the log--log fit to the matched sources, while the dashed line indicates the 1:1 (unity) relation. RASS-only and eRASS1-only detections are plotted along the respective axes. Factor-of-two dotted lines ($y = 2x$ and $y = x/2$) illustrate typical potential variability between the surveys. }
        \label{xrflux}
    \end{figure}

    The 41 RASS and eRASS1 common detections show a strong correlation in their X-ray fluxes with a Pearson correlation coefficient of $r = 0.905$ and a $p$ value of $\sim 5 \cdot 10^{-16}$ (see Fig. \ref{xrflux}), indicating a systematic relation between RASS and eRASS1 fluxes. In particular, the fluxes measured by eRASS1 scale consistently with RASS, suggesting that the observed differences are primarily due to instrumental sensitivity and calibration effects rather than intrinsic source variability. The best-fit line in log--log space has a slope of $0.82 \pm 0.06$ and an intercept of  $-2.2 \pm 0.7$. This slope is slightly below unity, a statistically significant difference, indicating a small systematic offset between RASS and eRASS1 flux scales. Most sources lie within a factor-of-two band around the 1:1 line ($y = 2x$ and $y = x/2$), suggesting that variability, such as flares, largely accounts for the differences. RASS-only and eRASS1-only sources are plotted along the axes to show unmatched fluxes. Nevertheless, given the strong correlation, this offset is unlikely to affect our overall results in any meaningful way. Based on the {\sc HamStar} algorithm, most of the stars had a coronal probability of ${p}_{\text{coronal}}>0.5$, indicating that they are likely coronally active stars.
    At optical wavelengths, the stars show a \textit{Gaia G}-band brightness range of 4.5 $<G<$ 17.8 magnitude, with ($BP-RP$) colours ranging from --0.19 to 5.05 mag.
    In the radio L band, the SMGPS's integrated flux density ranges from $0.05-170$ mJy.

    \subsection{Stellar populations \label{sec:pop}}

\begin{figure}[ht]
    \centering
        \includegraphics[width=\columnwidth]{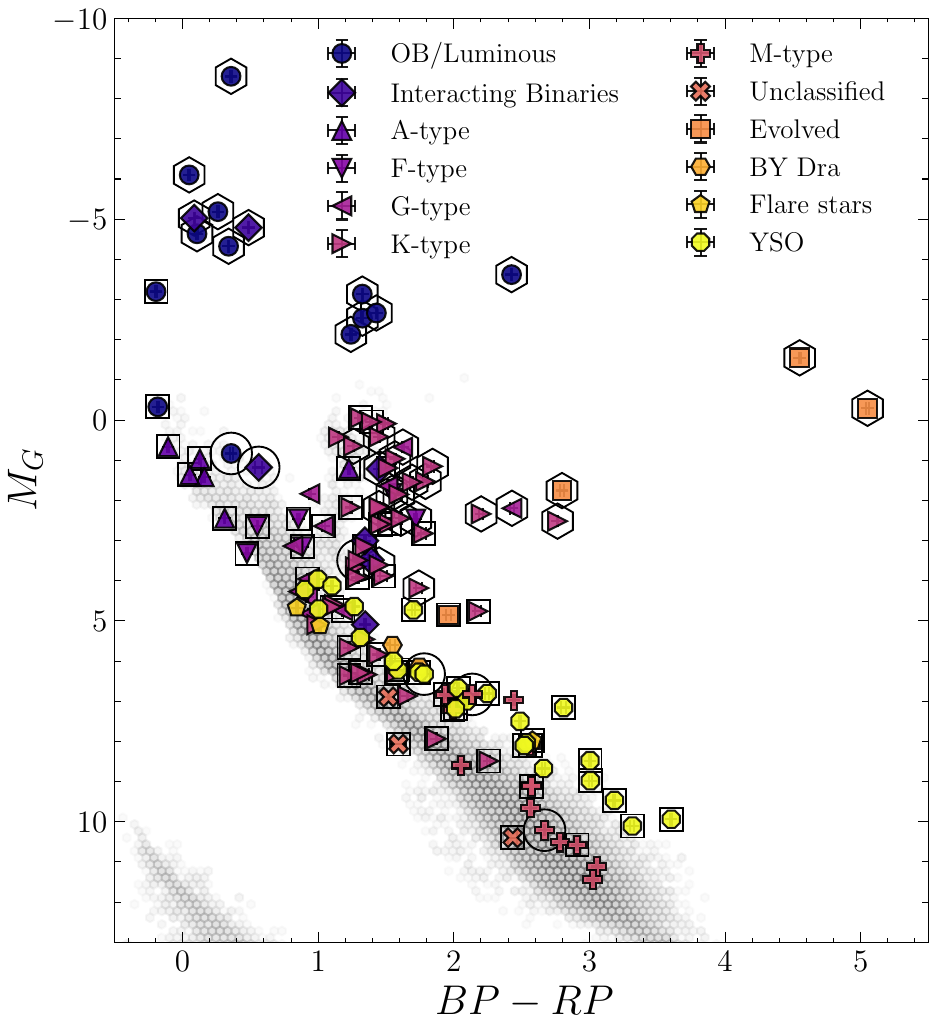}
        \caption{{\it Gaia} colour-magnitude diagram for all 137 RASS and eRASS1 radio- and X-ray-emitting stars. The markers represent stellar classifications reported in ({\tt SIMBAD}). The interacting binaries include the RS CVn, eclipsing, and spectroscopic binaries. Evolved stars comprise classical cepheids and long-period variables. The G-type, K-type, and M-type classifications are based on the {\it Gaia} astrophysical parameter catalogue \citep{GaiaCollaboration:2023}. The grey circles all represent {\it Gaia} DR3 sources within 100 pc of Earth, binned in uniform colour and magnitude. Large open circles represent the six detections in RASS that are not present in eRASS1, whereas the open squares show the 56 detections in eRASS1 that are absent in RASS and within $1 \, \rm kpc$; the open hexagon markers show the 34 eRASS1 stars farther than $1\, \rm kpc$ away. 
        }
        \label{cmd}
    \end{figure}

    All sources were cross-matched with the  {\tt SIMBAD} archive \citep{Wenger:2000} to identify source populations. Our sample contains many spectral types across the HR diagram, including main-sequence, pre-main-sequence, and evolved stars (mainly of classical Cepheids and long-period variables). Also included are single and binary systems (see Fig. \ref{cmd}). K- and M-type stars dominate the population, many of which are identified as young stellar objects (YSOs). There are a handful of interacting binaries comprising the RS CVn, eclipsing, and spectroscopic binary systems. Radio and X-ray emission are expected from these populations, though different mechanisms drive the emission. For early-type stars, free-free (bremsstrahlung) emission from ionised stellar winds is the dominant mechanism producing radio emission. X-ray emission arises from shocks within the stellar winds of luminous OB-type systems \citep{Guedel:1993, Gudel:2002, Gudel:2004}. For late-type stars, the radio and X-ray activity is driven by gyrosynchrotron and coronal emission, respectively \citep{Gudel:2002, Vedantham:2022, Driessen:2022}.

    The enhanced sensitivity of eRASS1 has significantly increased the number of known radio and X-ray emitting stars. While 41 stars are detected in both the RASS and eRASS1 samples, eRASS1 has enabled the additional detection of approximately three times as many stars, which were previously undetected in RASS.

    \begin{figure}[h!]
        \includegraphics[width=\columnwidth]{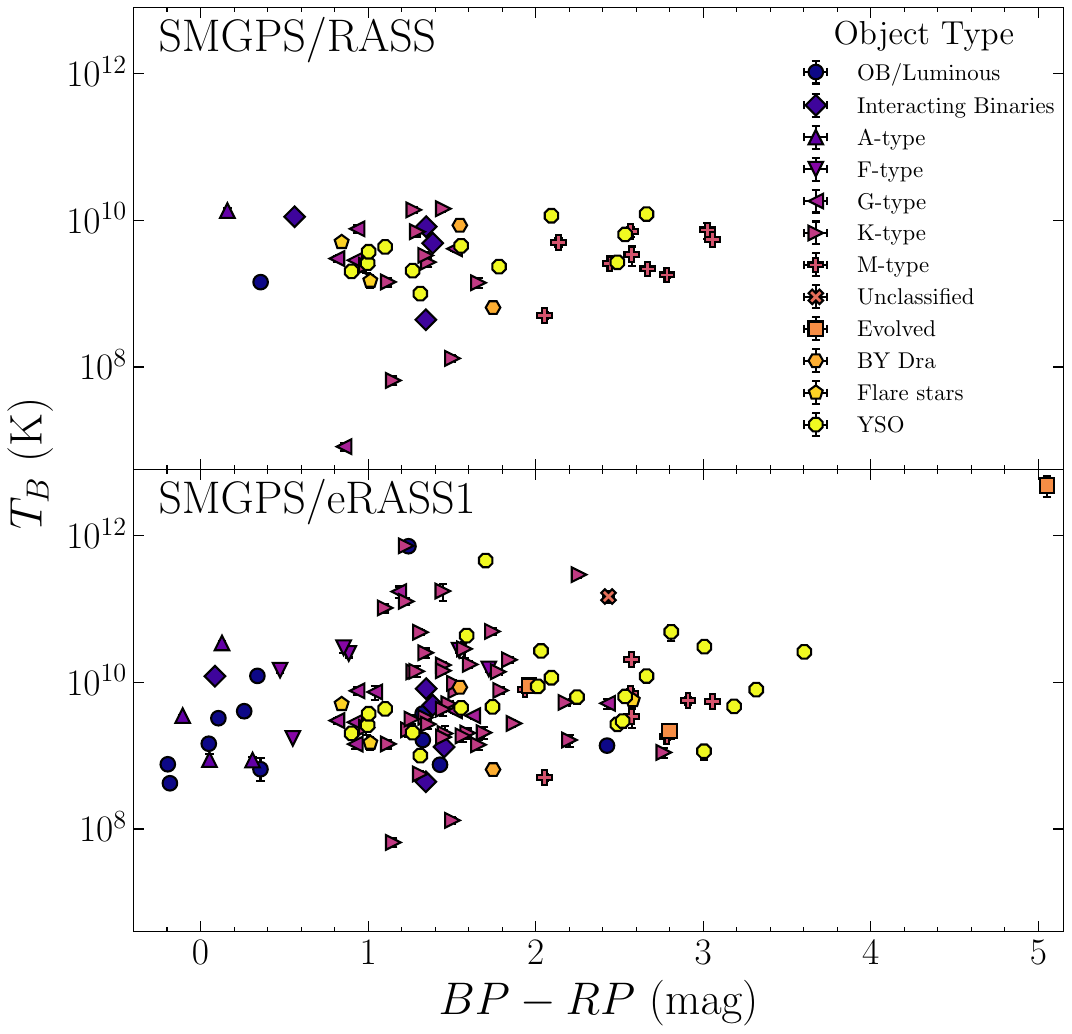}
         \caption{{\it Gaia} ($BP-RP$) colour versus brightness temperature, $T_{\rm B}$, for SMGPS--X-ray matches with RASS and eRASS1 detections. Markers represent stellar classifications reported in  {\tt SIMBAD}. The interacting binaries include the RS CVn, eclipsing, and spectroscopic binaries. Evolved stars comprise classical cepheids and long-period variables. The G-type, K-type, and M-type classifications are based on the {\it Gaia} astrophysical parameter catalogue \citep{GaiaCollaboration:2023}. The vertical error bars show the propagated uncertainties from the flux measurements and distances. 
         }
        \label{colbt}
    \end{figure}

    \subsection{Emission mechanisms}\label{emmech}
    
     We calculated the radio brightness temperature, $T_{\rm B}$, of all the detections as explained in Section \ref{btp}. We adopted the stellar radius provided by the Transiting Exoplanet Survey Satellite (TESS) input catalogue (TIC; \citealt{Stassun:2019, Paegert:2021}) for our sources. For sources with no radius value in the TIC, we searched the {\it Gaia} DR3 astrophysical parameters to obtain their radius. Also, we retrieved the \emph{Gaia} distance estimates provided by \cite{Bailer-Jones:2021}. We followed a similar approach to \cite{Pritchard:2021} in assuming that the emitting radius is three times the stellar radius ($R_e = 3R_*$). 

    The radio-brightness temperature, $T_B$, versus the \textit{Gaia} colour is given in Fig. \ref{colbt}, with the legend showing the spectral/stellar types. The radio-brightness temperatures vary among stars, which may reflect differences in coronal density, magnetic-field strength, and rotation rates.
    All our sources are at a brightness temperature of $T_{\rm B} <10^{12}\, \rm K,$ except for $\text{AX J1600.9--5142,}$ which has a brightness temperature of $T_{\rm B} = (4.8 \pm 1.5) \cdot 10^{12}\, \rm K$. 
    $\text{AX J1600.9--5142}$ is a well-known colliding-wind, double Wolf-Rayet binary, also known as \textit{Apep} \citep{delPalacio:2023}. An outlier at the lower end, with a $T_B$ of $(8 \pm 1) \cdot 10^{6}\, \mathrm{K}$, is HD 124831; it is a G3V-spectral-type binary system  \citep{Torres:2006}.

    The majority of the sources have radio-brightness temperatures in the range of $10^{9} - 10^{11}\, \mathrm{K}$. The eRASS1 sample includes sources brighter than $10^{10}$ K, unlike the RASS sample, whose maximum values are around $10^{10}$ K. 
    The sensitivity of eRASS1 enabled the identification of fainter and more distant stellar counterparts, reaching distances of up to $\sim 4500\, \rm pc$. Within the SMGPS/eRASS1 matched sample, 97 out of 131 sources lie within 1 kpc. 
    
    The radio-brightness temperature ranges obtained in this work are not uncommon for observations at a relatively low frequency of 1.3 GHz. For example, in \cite{Pritchard:2021}, all the stars are within the $\sim 10^{8} - 10^{11}$ K range, except for one late-type dwarf star with $T_B > 10^{12}$ K. Also, \cite{Ayanabha:2024} showed that most of the VLASS detections within 500 pc at 3 GHz are within $T_B < 10^{12}$ K, except for late spectral types beyond M2.5. 
    In this work, only $\text{AX J1600.9--5142}$ reached a brightness of $>10^{12}$ K, which is commonly adopted as the threshold distinguishing coherent from incoherent emission \citep{Dulk:1985, Gudel:2002, Pritchard:2021, Vedantham:2022, Ayanabha:2024}. Hence, we conclude that all remaining stars exhibit incoherent radio emission predominantly produced by gyrosynchrotron emission.

    \begin{figure}[h!]
        \includegraphics[width=\columnwidth]{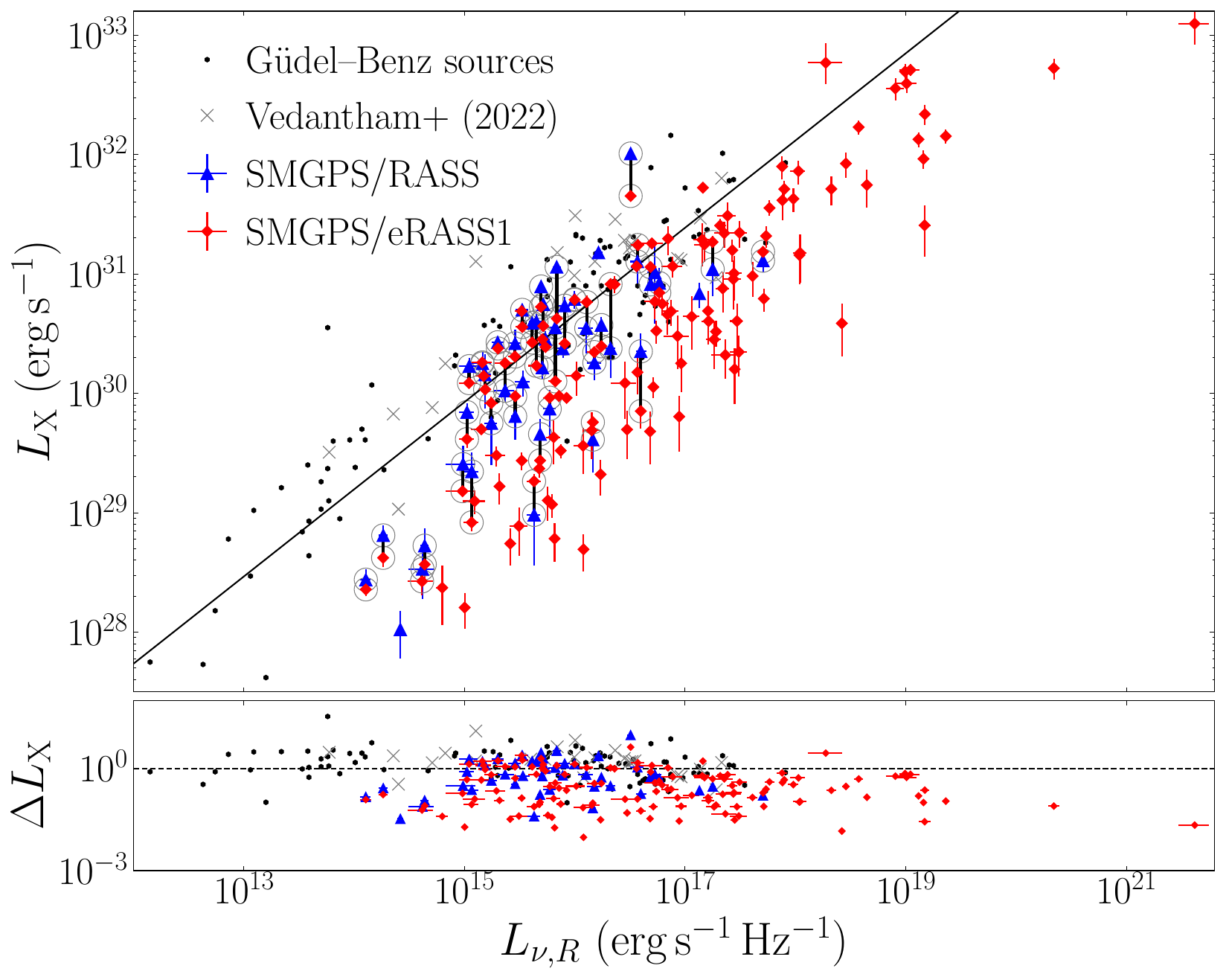}
        \caption{X-ray luminosity as function of specific radio luminosity for SMGPS, RASS, and eRASS1 stars. The triangles and diamonds are detections from RASS and eRASS1, respectively. The black dots and grey crosses are the data from \cite{Guedel:1993, Benz:1994} and \cite{Vedantham:2022}, respectively. The solid line in the top panel of the plot represents the G\"{u}del--Benz fit from \cite{Williams:2014}. The grey circles and solid vertical lines in the top panel indicate sources detected by both RASS and eRASS1. For each source, the vertical line connects the RASS and eRASS1 measurements, illustrating the level of X-ray variability. The bottom panel shows the residuals of the data from the GBR fit. The error bars in both panels indicate the propagated uncertainties in the measured fluxes and geometric distances for each source.
        }
        \label{gbenz}
    \end{figure}

    \subsection{Radio--X-ray-luminosity relationship}

    We investigated the GBR for our sources by calculating the radio and X-ray luminosities using the SMGPS radio fluxes and X-ray fluxes from both RASS and eRASS1, and geometric distance estimates information from \cite{Bailer-Jones:2021}. The radio and X-ray luminosity parameter space is shown in Fig. \ref{gbenz}, which also features the sources used to derive the canonical GBR from \cite{Guedel:1993} and \cite{Benz:1994}, as well as the {\sl LOFAR} sources from \cite{Vedantham:2022}. To enable a consistent comparison across samples, we note that the \cite{Guedel:1993, Benz:1994} and \cite{Vedantham:2022}, and our sample X-ray fluxes were derived using thermal coronal spectrum (e.g. Raymond--Smith, APEC) models appropriate for active stellar coronae \citep{Benz:1994, Schmitt:1995, Fleming:1995}.  
    For the radio data, the samples differ in observing frequency; the GBR used 5 -- 9 GHz measurements \citep{Benz:1994}, the SMGPS luminosities are based on 1.3 GHz flux densities, and the \cite{Vedantham:2022} sample used {\it LOFAR} observations at 144 MHz. Active stellar radio spectra are often flat or only mildly inverted in the gigahertz regime \citep[e.g.][]{Gudel:2002}, while low-frequency (100--200\,MHz) emission is dominated by non-thermal processes such as gyrosynchrotron or coherent bursts and still traces magnetic activity \citep{Zic:2019, Callingham:2021}. 
    These frequency differences may introduce some systematic offsets or scatter. Still, the comparison remains appropriate for assessing the overall form of the radio--X-ray correlation as the GBR is observed to hold over a range of frequencies from 200 MHz to 9 GHz. The solid line in Fig.\ \ref{gbenz} shows the canonical GBR fit parametrised by \cite{Williams:2014}. The RASS sources are scattered around the GBR fit with a dispersion of $\sim 0.53$ dex, whereas eRASS1 sources are scattered around the fit with a slightly larger dispersion of 0.57 dex. Overall, our samples are distributed along the GBR fit, but with a higher concentration of sources below the GBR fit line. The bottom panel plot in Fig. \ref{gbenz} shows the residuals from the GBR plotted against $L_{\nu,R}$ for both RASS and eRASS1 sources.

    In our sample, the canonical GBR appears to represent the upper envelope of the $L_{X} - L_{\nu, R}$ correlation. Stars with $L_X > 10^{29}\,\mathrm{erg\,s^{-1}}$ are broadly consistent with the GBR, although they exhibit a spread of more than 2 dex in $L_{\nu,R}$. There is also a notable absence of stars with $L_X$ between $10^{28}$ and $10^{29}\,\mathrm{erg\,s^{-1}}$ lying on the GBR; the detected objects in this regime show systematically higher $L_{\nu,R}$ than expected. Their X-ray luminosities fall within the range expected for coronally active stars, so the deviation cannot be explained by unusually weak X-ray emission. A plausible explanation is the radio sensitivity limit, whereby only stars observed during flaring or enhanced non-thermal states exceed the radio detection threshold, while quiescent emission remains undetected. However, we note that the small sky coverage of the survey ($\sim 1.2\%$) may also contribute to the observed distribution if such radio-bright states or objects are intrinsically rare.
    
    Specific radio luminosities above the canonical GBR are well established for several stellar regimes and can arise from sensitivity-driven selection effects in the radio band. For low-luminosity late-M and ultra-cool dwarfs, numerous studies have shown that radio emission is systematically over-luminous by several orders of magnitude, because only flaring or strongly magnetically active sources exceed current radio-detection thresholds for this population \citep{Berger:2006, Berger:2010, Williams:2014, Villadsen:2019}. Likewise, recent surveys of solar-type G and K stars show that most radio detections correspond to short-duration flares or enhanced non-thermal activity, while the quiescent levels remain below the instrumental sensitivity \citep{Yiu:2024, Driessen:2024}. These sensitivity effects can therefore broaden the observed $L_X$--$L_{\nu,R}$ distribution and shift the apparent offset to higher luminosities when sampling more distant or intrinsically brighter stellar populations. In the context of our sample, this implies that the excess radio luminosities we observe relative to the classical GBR may reflect the detection of only the most radio-active or magnetically active sources within the SMGPS survey, rather than a fundamental deviation in the underlying coronal physics.

    \begin{table*}
    \centering
    \caption{Power-law fits to luminosity relationships of{
    RASS} and {eRASS1} samples.
    }
    \label{tab:powerlaw}
    \begin{tabular}{lll}
    \hline
    Relation & RASS & eRASS1 \\
    \hline \\[0.1em]
    $L_{\rm X}$ -- $L_{\rm \nu,R}$ &
    $\displaystyle L_{\rm X} = 10^{17.2 \pm 1.6}  L_{\rm \nu,R}^{0.8 \pm 0.1}$ &
    $\displaystyle L_{\rm X} = 10^{19.1 \pm 0.6}  L_{\rm \nu,R}^{0.68 \pm 0.04}$ \\[1.2em]
    
    $L_{\nu,\mathrm{R}}$ -- $L_{\mathrm{bol}}$ &
    $\displaystyle L_{\rm \nu,R} = 10^{-2.1 \pm 2.5} L_{\rm bol}^{0.53 \pm 0.07}$ &
    $\displaystyle L_{\rm \nu,R} = 10^{-2.6 \pm 1.5} L_{\rm bol}^{0.56 \pm 0.04}$ \\[1.2em]

    $L_{\rm X}$ -- $L_{\rm bol}$ &
    $\displaystyle L_X = 10^{29.82 \pm 0.07} \left(\frac{L_{\rm bol}}{10^{33 }}\right)^{0.67 \pm 0.06}$ &
    $\displaystyle L_X = 10^{29.90 \pm 0.07} \left(\frac{L_{\rm bol}}{10^{33}}\right)^{0.48 \pm 0.03}$ 
    \\[0.5em]
    \hline
    \end{tabular}
    \label{lumtab}
    \end{table*}

    \begin{figure*}[h!]
        \sidecaption
        \includegraphics[width=12cm]{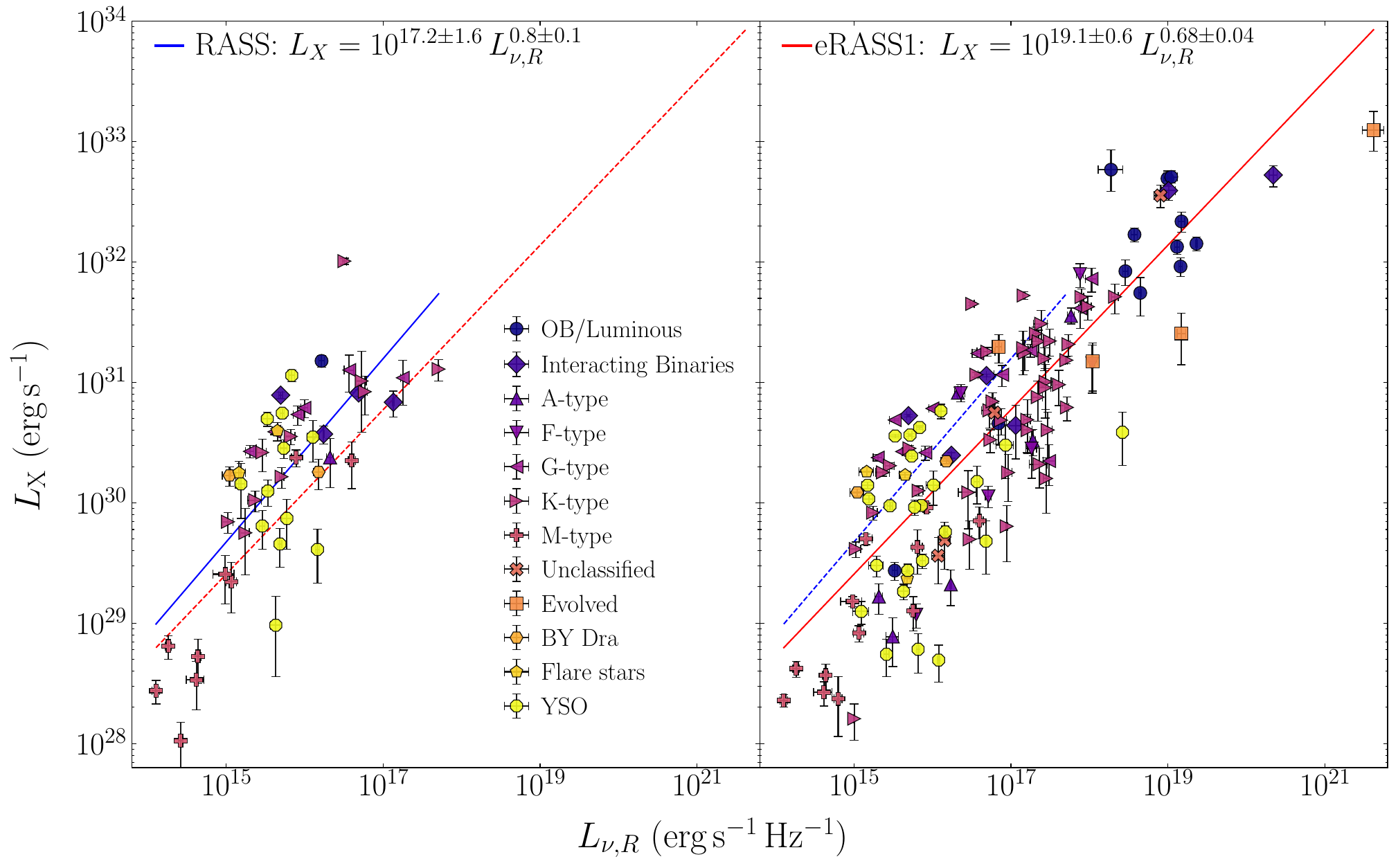}
        \caption{X-ray luminosity as function of specific radio luminosity for the SMGPS$-$RASS (left) and eRASS1 (right) sources. In each panel, the solid line shows the fit to that dataset, while the dashed line shows the fit obtained from the other dataset. The markers show object/spectral-type information obtained from either {\tt SIMBAD} or \cite{GaiaCollaboration:2023} {\it Gaia} DR3 astrophysical parameters.
        }
        \label{gbenz2}
    \end{figure*}

    \subsubsection{Power-law correlations}

    To assess whether our radio--X-ray-selected sample is consistent with the GBR, we fitted a power-law function in logarithmic space to the SMGPS specific luminosities and X-ray luminosities derived from RASS and eRASS1 counterparts, as shown in Fig. \ref{gbenz2}. Parameters of the best fits are given in Tab.\ \ref{tab:powerlaw}. The RASS--SMGPS correlation  shows a Kendall $\tau$ \citep{Kendall:1938} correlation coefficient of $\tau = 0.576$ with a $p$ value of $1.11\cdot 10^{-8}$, and the eRASS1--SMGPS correlation results in $\tau =0.626$ and a $p$ value of $1.20\cdot 10^{-29}$.

    The eRASS1-derived slope of the $L_X$--$L_{\nu,R}$ relation ($0.68 \pm 0.04$) is in close agreement with the \cite{Williams:2014} canonical GBR exponent of 0.73 (see Eq.\ \ref{gbreq2}), indicating that the fundamental scaling between coronal X-ray and non-thermal radio emission remains valid for our sample. The fitted normalisation is within $2\sigma$ of the canonical relation with no quoted uncertainty.
    On the other hand, the RASS relation yields a slightly steeper slope ($0.8 \pm 0.1$), but with a significantly larger scatter in the $L_{\rm X}$--$L_{\rm R}$ plane. The broad dispersion observed for RASS may reflect a combination of factors, including {\sl ROSAT}'s lower sensitivity level, which may bias detections towards variable or X-ray-luminous subsets of sources. 

    Generally, both RASS and eRASS1 have roughly the same vertical spread across the GBR. However, eRASS1 allows for the detection of objects approximately one order of magnitude more luminous than the brightest RASS source in our sample, with the SMGPS radio luminosity being three orders of magnitude brighter than that of the RASS samples. Comparing our SMGPS samples to GBR, some GBR sources appear less luminous in both the radio and X-rays. This can be attributed to the flux limit of our sample. In addition, two stars (AX J1600.9--5142 and HD 93129A) show excessive radio luminosities beyond $10^{20} \, \mathrm{erg \, s^{-1}\, Hz^{-1}}$. Both systems are very-early-type interacting binaries \citep{delPalacio:2023,Benaglia:2025}.

    By contrast, {\it eROSITA}'s improved sensitivity provides a more complete sampling of coronal X-ray emission, especially among fainter sources that were below the detection limit of {\it ROSAT}. This yields a tighter $L_X$--$L_{\nu,R}$ correlation for the coronal population. Although the eRASS1 sample extends to larger distances and contains a broad mix of stellar types, the tighter relation arises because {\it eROSITA} detects the full range of coronal luminosities rather than only the brightest or most variable objects. Non-coronal emitters (e.g. massive OB stars and interacting binaries, whose X-rays are wind driven) do not dominate the fitted correlation \citep{Freund:2024}, ensuring that the observed trend reflects the intrinsic behaviour of magnetically active coronal sources rather than contamination from unrelated emission mechanisms.

    \subsubsection{Implications for coronal emission and stellar populations}
    
    Our findings highlight the importance of sample selection and instrument sensitivity in characterising correlations such as the GBR. While both RASS and eRASS1 samples show broadly consistent scaling behaviour, the tighter relation obtained with eRASS1 supports its robustness across a more diverse sample with higher statistics. 

    \begin{figure*}[h!]
        \sidecaption
        \includegraphics[width=12cm]{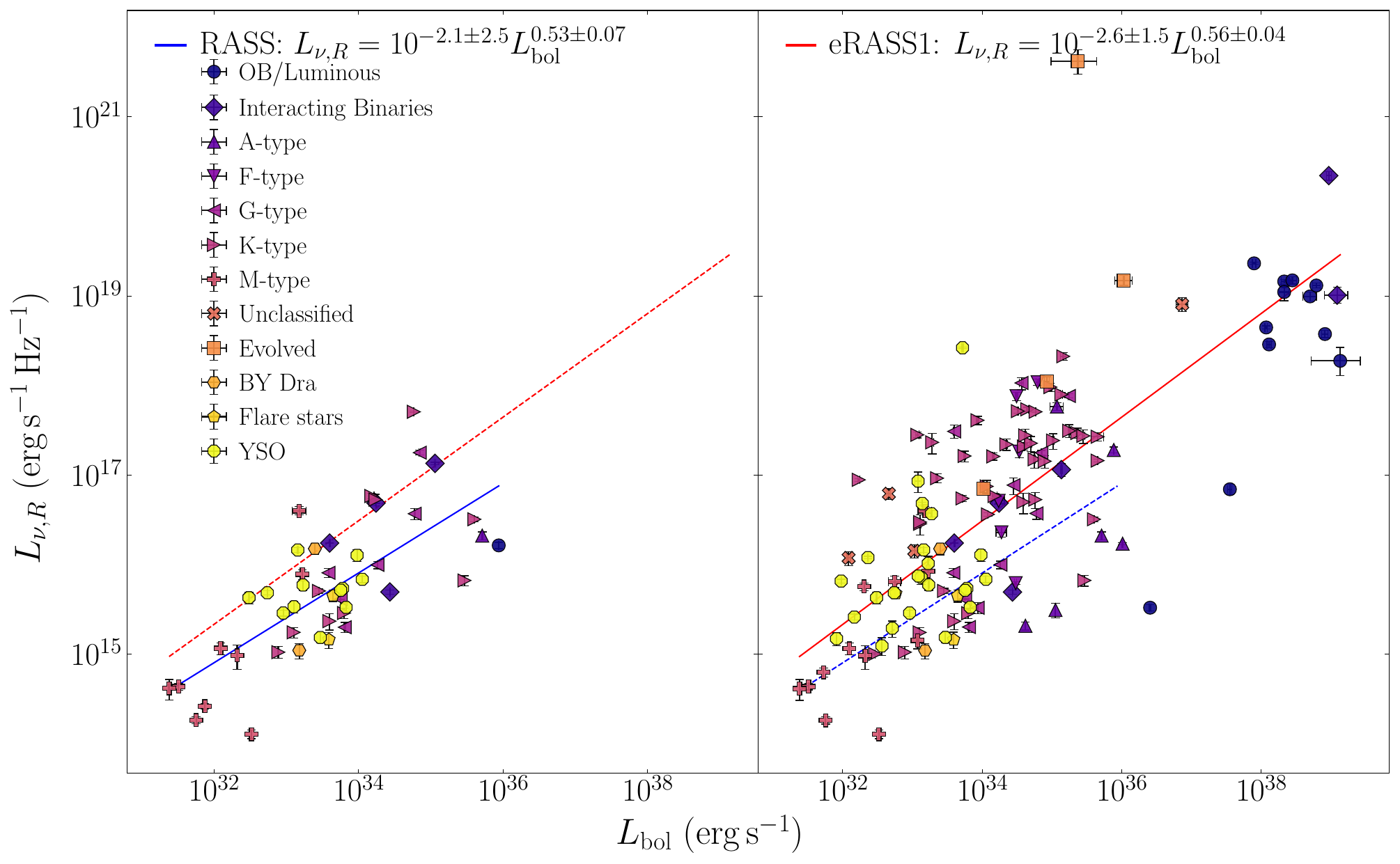}
        \caption{Specific radio luminosity, $L_{\rm \nu,R}$, as a function of bolometric luminosity, $L_{bol}$, for the SMGPS$-${RASS, eRASS1} sources. The solid and dashed lines, and markers are the same as in Fig. \ref{gbenz2}. The blue and red lines in both panels are the power-law fit for {RASS} and {eRASS1}, respectively.
        }
        \label{optrad}
    \end{figure*}

    \subsection{Comparison with optical luminosities}
\begin{figure*}
        \sidecaption
        \includegraphics[width=12cm]{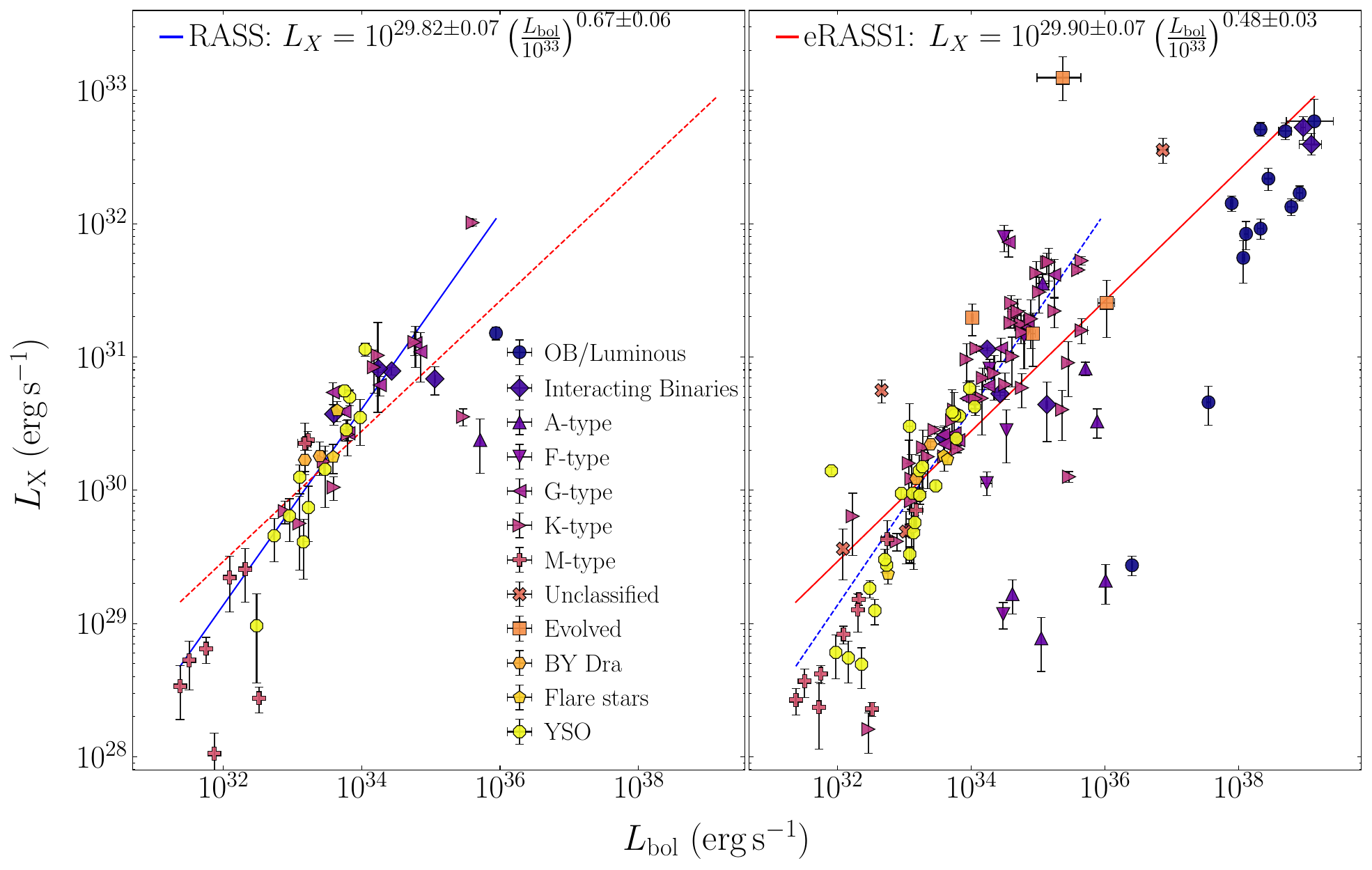}
        \caption{X-ray luminosity, $L_{\rm X}$ as a function of bolometric luminosity, $L_{\rm bol}$ for the SMGPS$-${RASS, eRASS1} sources. The solid and dashed lines, and markers are the same as in Fig. \ref{gbenz2}. The blue and red lines in both panels are the power-law fit for {RASS} and {eRASS1}, respectively.
        }
        \label{optxr}
    \end{figure*}

    Using the identified samples, we examine correlations of $L_{\nu,R}$ and $L_{\rm X}$ with the bolometric\footnote{We computed $L_{\rm bol}$ from the de-reddened \textit{Gaia} $G$-band magnitude, applying bolometric corrections based on spectral classes from a table based on \cite{Pecaut:2013}, which is regularly updated at \url{http://www.pas.rochester.edu/~emamajek/EEM_dwarf_UBVIJHK_colors_Teff.txt} (current version 2022.04.16).} luminosity, $L_{\rm bol}$. We fitted the correlations with a power-law function and show the relationships for all cases in Table \ref{lumtab} and Figs \ref{optrad} and \ref{optxr}. In Fig. \ref{optrad}, both the RASS and eRASS1 samples show no obvious correlation along the best-fit relation. However, for eRASS1, we see two clusters of populations: one below $L_{\rm bol} \sim 10^{37}\,\rm erg\,s^{-1}$ dominated by coronally active stars; and above it, the massive high-luminosity sources whose radio-emission origins are driven by thermal wind.  In Fig.~\ref{optxr}, the $L_X$--$L_{\rm bol}$ relation shows a clear correlation for the RASS sample, with sources broadly scattered around the best-fit line. In contrast, the eRASS1 distribution becomes highly irregular at higher bolometric luminosities, showing no clear correlation. When restricting the comparison to coronal sources only, the RASS-derived fit provides a better description of the coronal subset within eRASS1, whereas the non-coronal eRASS1 objects (principally OB and evolved stars) deviate strongly from any $L_X$--$L_{\rm bol}$ trend. The broader diversity seen in Figs. \ref{optrad} and \ref{optxr} is revealed only by the higher sensitivity of eRASS1, where RASS does not reach into the domain where the correlation between $L_{\rm X}\, \mathrm{versus} \, L_{\rm bol}$ and $L_{\rm \nu,R}\, \mathrm{versus} \, L_{\rm bol}$ breaks down. 

\begin{figure}[h!]
    \centering
      \includegraphics[width=\columnwidth]{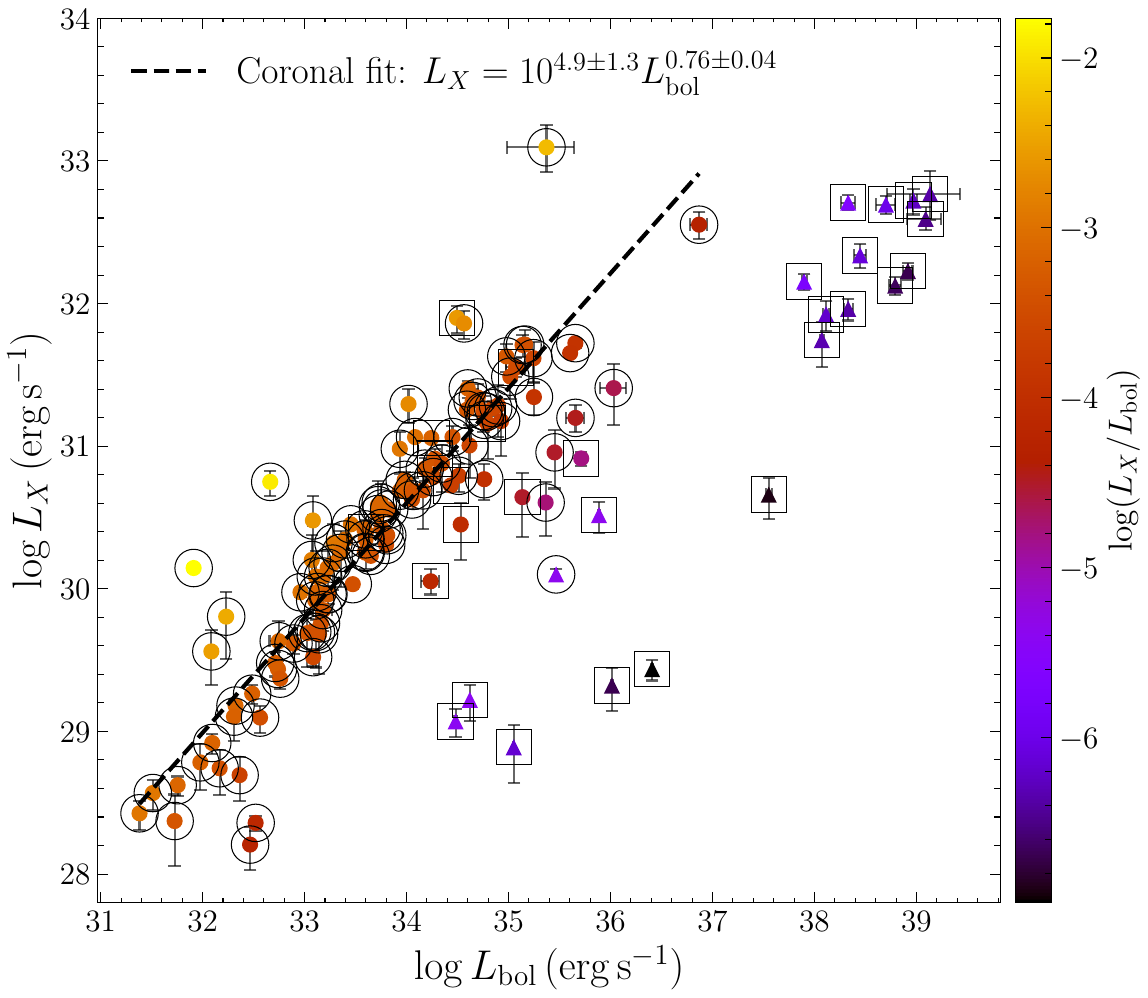}
        \caption{X-ray luminosity, $L_X$, as a function of bolometric luminosity, $L_{\rm bol}$, for the eRASS1-detected stars colour coded by the $\log (L_X/L_{\mathrm{bol}})$. Circles and triangles indicate sources with $\log (L_X/L_{\mathrm{bol}}) \geq -5$ and $\log (L_X/L_{\mathrm{bol}}) < -5$, respectively. In our flux-limited sample, the former group is dominated by coronally active stars, while the latter is consistent with non-coronal (wind-driven) emitters. The dashed line represents the linear fit in log-log space for all the stars (indicated with large open circles), excluding OB-luminous, A-type, F-type and intermediate binaries, which are indicated with open squares.
        }
        \label{corowind}
    \end{figure}

    Given that the early-type stars detected in eRASS1 do not conform well with the $L_{\rm X}$ vs $L_{\rm bol}$ correlation, we excluded the early-type samples comprising OB-luminous, A-type, F-type, and intermediate binaries from further analysis. Specifically, we focused on fitting the X-ray luminosity as a function of bolometric luminosity, $L_{\rm bol}$, for the corona-dominated later type stars in contrast to the full-sample fits presented in Table \ref{lumtab}; this was done to constrain the coronal emission component and obtain a more meaningful fit.
    We achieved a linear fit of $L_{\rm X} = 10^{4.9 \pm 1.3} L_{\rm bol}^{0.76 \pm 0.04}$ (see Fig. \ref{corowind}), indicating that the X-ray luminosity, $L_{\rm X}$, scales with bolometric luminosity, $L_{\rm bol}$, with an exponent of $0.76$; this indicates that the growth of $L_{\rm X}$ with $L_{\rm bol}$ is shallower than a one-to-one relation. In addition, most stars in this sample have $\log (L_{\rm X}/L_{\rm bol}) \approx -3$, with a few reaching values as low as $-7$.  The clustering around $-3$ is consistent with the coronal saturation limit observed in rapidly rotating, magnetically active stars, while the lower ratios likely represent less active and older coronae. This sample separation in Fig. \ref{corowind}, where $\log (L_X/L_{\mathrm{bol}}) \geq -5$ represents coronal sources and $\log (L_X/L_{\mathrm{bol}}) < -5$ represents non-coronal (wind-driven) emitters, is not a universal boundary, as many coronal stars, including the Sun with $\log(L_X/L_{\mathrm{bol}}) \approx -6$, fall below this threshold. The apparent separation arises because eRASS1 preferentially detects the more luminous coronal sources (near the saturation limit $\log(L_X/L_{\mathrm{bol}}) \sim -3$), while lower activity coronae can only be observed within a small local volume. The slope below unity implies that as $L_{\rm bol}$ increases, the relative X-ray output ($L_{\rm X}/L_{\rm bol}$) decreases. This trend is more consistent with reduced dynamo efficiency in more luminous, more slowly rotating stars. This behaviour is well established in studies of coronal X-ray scaling (e.g. \cite{Gudel:2004} and \cite{Reiners:2014}) and supports the interpretation that late-type stars in our sample are dominated by magnetically heated stellar coronae.

\section{Conclusions}

    We identified 137 radio and X-ray emitting stars obtained by cross-correlating the SARAO MeerKAT Galactic-plane compact source catalogue with soft X-ray detections from RASS and eRASS1. The radio-brightness temperature of almost all the stars in the sample is $\leq10^{12}$ K, indicating that the radio detections at 1.3 GHz are consistent with incoherent gyrosynchrotron emission from mildly relativistic electrons in stellar coronae, as observed in other magnetically active stars. We find that the radio and X-ray luminosities of our sample are correlated; however, the GBR does not strictly hold for all sources. Instead, it appears to trace the upper envelope of the $L_{\rm X} - L_{\nu,R}$  distribution, suggesting that while some stars approach the canonical relation, many lie below it. When compared to the canonical GBR, the majority of sources observed at 1.3 GHz appear over-luminous in the radio relative to the canonical GBR. This offset is primarily driven by enhanced radio luminosities rather than suppressed X-ray emission, which is consistent with the broader radio luminosity distribution sampled by the eRASS1 population.  
    Our power-law fits to the $L_{\rm X}$--$L_{\rm \nu,R}$ relation yield slopes consistent with those of the canonical GBR for both RASS and eRASS1, indicating that the relative scaling between the coronal soft X-ray and non-thermal radio emission is similar for our population, but the normalisation differs. Additionally, the eRASS1 sources reveal a more diverse population, including massive/luminous sources, compared to that based on RASS. We derived scaling relations for specific radio and bolometric luminosities ($L_{\rm \nu,R}$ versus $L_{\rm bol}$) and X-ray and bolometric luminosities $(L_{\rm X}$ versus $L_{\rm bol})$. These relations reveal trends in our data that may reflect underlying stellar properties, such as rotation and spectral type, thanks to eRASS1 data. In examining the $L_{\rm X}$--$L_{\rm bol}$ relation, we found that the early-type stars do not follow the same trend as the late-type stars. By fitting only the late-type stars, we obtained a robust power-law fit for the X-ray-versus-bolometric luminosity with most stars showing $\log (L_{\rm X}/L_{\rm bol}) \sim -3$, which is consistent with coronal saturation levels.

    Looking forward, future \textit{eROSITA} all-sky survey data releases (DR2 and DR3) will provide deeper X-ray coverage and improved statistics for stellar samples, enabling more robust population studies. At the same time, next-generation radio facilities, most particularly the Square Kilometre Array (SKA), will deliver unprecedented sensitivity levels. Together, the datasets will provide powerful synergies for probing stellar magnetic activity and refining our understanding of radio- and X-ray-emitting stars in our Galaxy.

\section*{Data availability}

    Information about the SARAO Galactic Plane Survey can be found in \cite{Goedhart:2024} and the compact source catalogue in \cite{Mutale:2025}. The coronal X-ray data from ROSAT and eROSITA eRASS1 are available in \cite{Freund:2022} and \cite{Freund:2024}, respectively.
    The Gaia DR3 data used in this work are available in the CDS Vizier database; the details can be found in \cite{ GaiaCollaboration:2023}. The radio-X-ray star catalogue from this work will be made available at the CDS (Centre de Donn\'ees astronomiques de Strasbourg) via anonymous ftp to \href{http://cdsarc.u-strasbg.fr}{cdsarc.u-strasbg.fr (130.79.128.5)}  or via \href{http://cdsweb.u-strasbg.fr/cgi-bin/qcat?J/A+A/}{http://cdsweb.u-strasbg.fr/cgi-bin/qcat?J/A+A/} upon publication, and the details of the column description are listed in \ref{tab:cols}

\begin{acknowledgements}

    The authors would like to thank H. K. Vedantham for sharing his code and data for making the radio--X-ray luminosity plot.

    This work is based on research fully supported by the National Research Foundation's (NRF) South African Astronomical Observatory (SAAO) PhD Prize Scholarship. PJG and ODE are (partly) supported by the NRF SARChI grant 111692. DAHB acknowledges research support by the National Research Foundation. JR acknowledges support from DLR under 50QR2505.

    The MeerKAT telescope is operated by the South African Radio Astronomy Observatory, which is a facility of the National Research Foundation, an agency of the Department of Science and Innovation.

    We acknowledge the use of the ilifu cloud computing facility -- \href{www.ilifu.ac.za}{www.ilifu.ac.za}, a partnership between the University of Cape Town, the University of the Western Cape, Stellenbosch University, Sol Plaatje University and the Cape Peninsula University of Technology. The ilifu facility is supported by contributions from the Inter-University Institute for Data Intensive Astronomy (IDIA -- a partnership between the University of Cape Town, the University of Pretoria and the University of the Western Cape), the Computational Biology division at UCT and the Data Intensive Research Initiative of South Africa (DIRISA).

    This work has made use of data from the European Space Agency (ESA) mission \href{https://www.cosmos.esa.int/Gaia}{\it Gaia}, processed by the {\it Gaia} Data Processing and Analysis Consortium \href{https://www.cosmos.esa.int/web/Gaia/dpac/consortium}{(DPAC)}. Funding for the DPAC has been provided by national institutions, in particular the institutions participating in the {\it Gaia} Multilateral Agreement. 

    This research has made use of the VizieR catalogue access tool, CDS, Strasbourg, France (\href{DOI:10.26093/cds/vizier}{DOI:10.26093/cds/vizier}). The original description of the VizieR service was published in 2000, A\&AS 143, 23.

    This work is based on data from eROSITA, the soft X-ray instrument aboard SRG, a joint Russian-German science mission supported by the Russian Space Agency (Roskosmos), in the interests of the Russian Academy of Sciences represented by its Space Research Institute (IKI), and the Deutsches Zentrum f\"ur Luft- und Raumfahrt (DLR). The SRG spacecraft was built by Lavochkin Association (NPOL) and its subcontractors, and is operated by NPOL with support from the Max Planck Institute for Extraterrestrial Physics (MPE). The development and construction of the eROSITA X-ray instrument was led by MPE, with contributions from the Dr. Karl Remeis Observatory Bamberg \& ECAP (FAU Erlangen-Nuernberg), the University of Hamburg Observatory, the Leibniz Institute for Astrophysics Potsdam (AIP), and the Institute for Astronomy and Astrophysics of the University of T\"ubingen, with the support of DLR and the Max Planck Society. The Argelander Institute for Astronomy of the University of Bonn and the Ludwig Maximilians Universit\"at Munich also participated in the science preparation for eROSITA. The eROSITA data shown here were processed using the eSASS software system developed by the German eROSITA consortium.

    This research used \href{http://www.astropy.org}{\tt ASTROPY}, a community-developed core Python package for Astronomy \citep{AstropyCollaboration:2013, AstropyCollaboration:2018}. We also use {\tt NUMPY} \cite{Harris2020}, {\tt SCIPY} \citep{Virtanen2020}, {\tt PANDAS} \citep{reback2020pandas},  and {\tt MATPLOTLIB} \citep{Hunter:2007}. This research has used the SIMBAD database, operated at CDS, Strasbourg, France.

\end{acknowledgements}

\bibliographystyle{aa}
\bibliography{biblio}

\begin{appendix} 
\section{Column description}
\begin{table}[] 
\centering
\caption{Column definitions with units and descriptions.}\label{tab:cols}
\begin{tabular}{lll}
\hline
\textbf{Column} & \textbf{Unit} & \textbf{Description} \\
\hline
SMGPS\_id & -- & SMGPS Unique source identifier \\
fileName & -- & Name of the FITS or data file \\
lon & deg & Source Galactic longitude \\
lat & deg & Source Galactic latitude \\
ra & deg & Source right ascension (J2019.4) \\
dec & deg & Source declination (J2019.4) \\
ra\_err & arcsec & Uncertainty in right ascension \\
dec\_err & arcsec & Uncertainty in declination \\
radec\_err & arcsec & Error in both right ascension and declination in arcsec \\
peak\_flux & mJy/beam & Maximum measured flux density \\
err\_peak\_flux & mJy/beam & Uncertainty in peak flux \\
int\_flux & mJy & Source integrated  flux density \\
err\_int\_flux & mJy & Uncertainty in source integrated flux \\
ERO/2RXS & -- & IAU name of the eRASS1/RASS source (IAUNAME) \\
FX & erg s$^{-1}$ cm$^{-2}$ & X-ray flux in the 0.2 -- 2.3 keV band assuming a coronal spectrum \\
FX\_ujy & $\mu$Jy & X-ray flux converted to $\mu$Jy units \\
p-coronal & -- & Probability of the eRASS1 source to be coronal \\
Source & -- & Gaia DR3 source identifier \\
RA\_ICRS & deg & Gaia right ascension in ICRS (J2016.0) \\
DE\_ICRS & deg & Gaia declination in ICRS (J2016.0) \\
Gmag & mag & Gaia broad-band $G$ magnitude \\
BP-RP & mag & Gaia colour index (BP - RP) \\
rgeo & pc & Gaia geometric distance estimate \\
mgg & mag & Absolute $G$ magnitude \\
b\_temp & K &  Stellar radio brightness temperature \\
r\_lum & erg$\,$s$^{-1}\,$Hz$^{-1}$ & Computed specific radio luminosity at 1.3 GHz \\
x\_lum & erg s$^{-1}$ & Computed X-ray luminosity \\
Lerg & erg s$^{-1}$ & Bolometric luminosity \\
main\_id & -- & Main identifier from Simbad \\
otype\_lbl & -- & Object type label from Simbad \\
main\_type & -- & Object type from Simbad \\
SpType-ELS & -- & Stellar spectral type from Gaia DR3 if available \\
type\_note & -- & Additional notes on classification \\
sep\_arcsec & arcsec & SMGPS- Gaia separation in arcsec \\
\hline
\end{tabular}
\end{table}

\end{appendix}

\end{document}